# A wireless passive pressure sensor with high sensitivity

Baiyun Wang[1], Yijia Cheng[2], Yujie Hua[1], Wenxuan Tang[*1], Wei E.I. Sha[2], Hao Xu[1], Kang Wang[3], Jundi Hu[3], Huaqing Fan[1], Huanran Peng[1], Gang Shao[3]

1 State Key Laboratory of Millimeter Waves, Southeast University, 210096 Nanjing, China
2 College of Information Science and Engineering, Zhejiang University, Hangzhou 310027, China
3 School of Materials Science and Engineering, Zhengzhou University, Zhengzhou 450001, China

## ➤ Abstract

A high-sensitivity wireless pressure sensor with active processing structure designed on the dielectric substrate has been present and evaluated in this paper. The sensor configuration has been optimized by computer-aided design to achieve highest sensitivity and maximum working range for a given dimension. With the average sensitivity of 187kHz/kPa, the proposed pressure sensor is equipped with the ability to measure pressure loaded up to 1.5MPa under room temperature. Additionally, a novel simulation method applied on pressure related design is proposed in this article, with the accuracy reaching threefold enhancement, filling the blank of electromagnetic simulation of pressure deformation. Other characteristics of the devices have been investigated and are presented.

## ➤ Introduction

Pressure is an important quantity and measured physical property in various fields, such as manufacturing, petrochemicals, and aerospace industry[1-4]. In aerospace field, various applications necessitate the handling of high-pressure scenarios, and miniaturized sensor with high sensitivity is the trend of development[5]. Researchers and academics have begun to search for new ways to ensure the sensing accuracy while reducing the size of sensors, such as employing integrated circuit fabrication technology to develop magnetic sensors[6].

To date, pressure sensors widely reported including piezoresistive pressure sensor, capacitive pressure transducers, piezoelectric pressure sensors[7], and optical pressure sensor are widely used in the aerospace industry[8]. Among them, piezoelectric and piezoresistive pressure sensor have been adopted on MEMS instruments and have reached consumers[9]. Despite of the simplicity of fabrication on suitably designed diaphragm[10], piezoresistive pressure sensors are sensitive to temperature variations[11] and suffer from lower repeatability[12] which is not an appropriate principle for pressure sensor design. While optical pressure sensors offer high sensitivity for low-pressure sensing[13], their drawback stems from the narrow range of measurement and the heavy reliance on light.

When considering high-pressure application, several common design requirements should be identified, including wireless telemetry, high reliability and batch fabrication[14]. In monitoring systems, sensors with wireless technique are vital in many applications due to their flexibility and scalability, enabling continuous monitoring of pressure changes without physical connections[15].

They offer a mobile and adaptable solution, eliminating the constraints of wired installations in impractical environments.

Taking leverage of the achievements in materials and manufacturing technologies, the combination of multiple pressure sensing principles becomes a good way of improving the accuracy and guaranteeing high sensitivity of sensors. Among the above measuring techniques, pressure sensors based on LC resonance principle with a deformable cavity has attracted more researchers for further exploration. Due to its notable advantages such as high sensitivity, broad measurement range and passive operation, LC resonance sensors can exhibit good performance across various applications, especially ensure robust performance against external environmental factors, contributing to measurement stability and accuracy. Most LC resonance pressure sensors use a diaphragm that deflects in response to pressure change so that the pressure can be obtained indirectly by measuring the related parameters of sensors.

The simplest form of passive telemetry implements an LC circuit whose resonant frequency varies according to some environmental change. Passive LC pressure sensor was firstly proposed by Collins as early as 1967 to realize pressure monitoring in eyes[6]. With a pair of spiral coils constituting a distributed resonant circuit, the resonant frequency can shift linearly as the diaphragms of sealed cavity deflects by pressure. Based on the similar structure and principle, a completely passive wireless device without the need for contacts, feedthrough wires, or internal power supplies was proposed by English and Allen in late 1990s[16]. Since then, many low-profile passive wireless pressure sensors have been developed [17-21].

Polytetrafluoroethylene (PTFE) laminates are commonly used for RF/microwave circuits due to their excellent electrical performance at higher frequencies[22]. Rogers RO4003C materials are proprietary woven glass reinforced hydrocarbon/ceramics with the electrical performance of PTFE/woven glass and the manufacturability of epoxy/ glass[23], which can be used as a good material in radio frequency circuit. Apart from its good electrical performance, RO4003C also demonstrates favorable mechanical properties suitable for applications involving pressure. It boasts a Yang's modulus of 19.65 GPa and a Poisson's ratio of 0.2. Most FR-4 prepreg materials exhibit strong bonding characteristics with a PTFE laminate. This is attributed to the ability of the PTFE substrate surface to retain a mirror image of the etched copper, creating a textured surface that facilitates effective adhesion between the PTFE and FR-4 prepreg.

In this paper, we explore a wireless pressure sensor with high sensitivity and a wide measurement range up to 1.5MPa. Differing from previous research, the proposed sensor features a novel resonant structure based on LC resonance principle, utilizing designed metallic structure fabricated with Rogers 4003C materials. To enhance performance, we implemented an array-arranged LC model featuring capacitance and inductance, aiming to achieve both high Q factor and electrical compactness. To ensure the airtightness, FR4 material is used as prepreg to maintain a stable and well-defined air cavity shape. Furthermore, a new approach to simulate diaphragm deformation in pressure sensor is proposed for precise simulation.

## ➢ The design of single circular patch LC circuit

The design of the pressure sensor involves the creation of flexible membrane, the sealed cavity and the integration of the LC resonant circuit. To enhance the sensor's performance, the focus of this section is on designing the electromagnetic resonance structure. The schematic diagram of a single circular patch LC sensor working at X band is presented in Fig.1. The substrate of the sensor consists of three layers (the top layer, the bottom layer and the air cavity layer), wherein the two metal layers are separately attached on both sides of the substrate. As shown in Fig.1(b), the metallic circular structure on the surface is connected with the bottom metal ground by four metal strips and four vias on the four sides of the substrate. Rogers 4003C is used as the substrate, and the thickness of the three layers are 0.813mm, 0.305mm, 0.813mm respectively. The vital parameters of the design and the corresponding value are presented in Table 1.

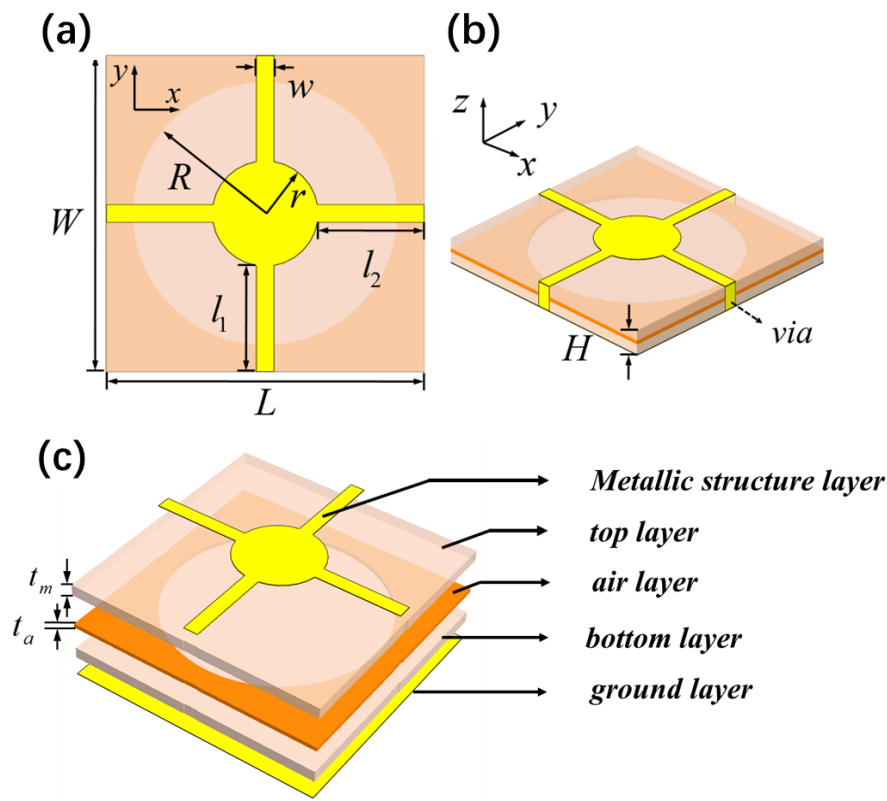

Figure 1 The schematic view of single circular patch LC circuit.

Table 1 the vital parameters of the single patch LC model

| Parameters | $r$ | $w$ | $l_1$ | $l_2$ | $R$ | $W$ | $L$ | $H$ |
|---|---|---|---|---|---|---|---|---|
| Value (mm) | 3 | 2.1 | 6 | 6 | 7.5 | 18 | 18 | 1.931 |

## ✦ The operation principle

### ➢ The Electromagnetic Analysis

Utilizing the electromagnetic simulation software CST Microwave Studio, full-wave simulations are conducted for both TE mode and TM mode. In this setup, the boundaries in the $x$ and $y$ directions are defined as the unit cell, while those in the z-direction are configured as open. The Frequency Domain Solver in CST, renowned for its efficacy in scrutinizing and refining resonance frequencies in electromagnetic structures, is selected as the instrumental tool to accurately calculate the resonance frequency for the pressure sensor.

Fig.2 depicts the electric field distribution and surface currents in TE mode. In Fig.2(a), the color gradient represents the electric field intensity, revealing that the predominant energy is concentrated around the circular plate in the center, with some radiation extending towards the vertical metal strip lines. Since the sensor's backside comprises a solid metal ground, the energy stored in the capacitor primarily accumulates at the plate's edge. Fig. 2(b) illustrates that the surface currents are concentrated at the vertical strip lines, suggesting that these lines can be regarded as inductors in the equivalent circuit. Simultaneously, the energy distributes near the metallic structure, encompassing the air cavity.

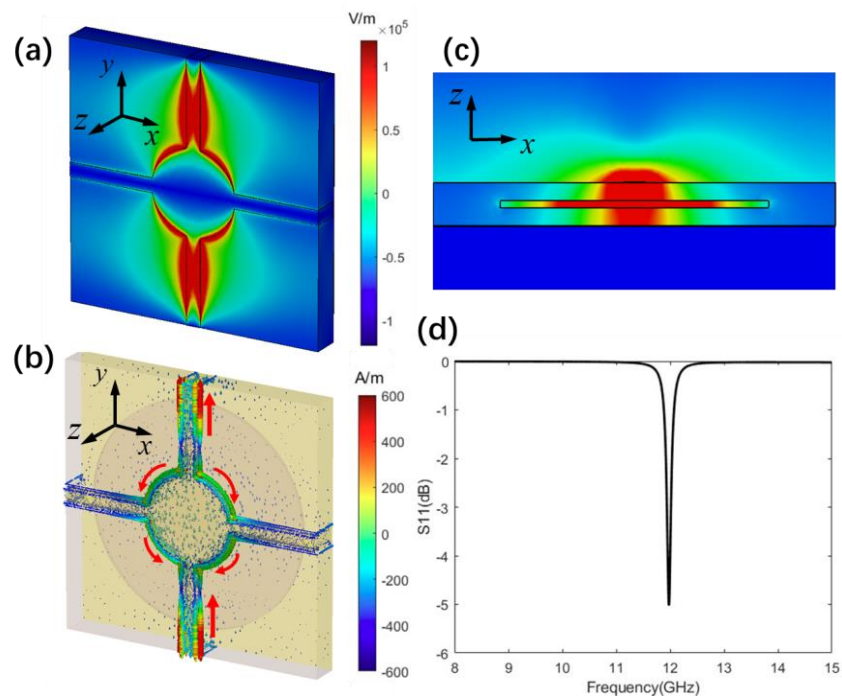

Figure 2 The simulated result of a single circular patch LC circuit working in TE mode. (a)Electric field at resonant frequency. (b) The distribution of surface current at resonant frequency. (d) Electric field from side sectional view at resonant frequency. (c) Simulated reflection coefficient (S11) parameter in CST.

The single patch pressure sensor presented in this work is based on the principle of LC resonant circuit model. For a LC resonant circuit, it demonstrates a distinctive response characterized by heightened sensitivity at a particular frequency, commonly referred to as the resonance frequency

$$f = \frac{1}{2\pi\sqrt{LC}}.$$

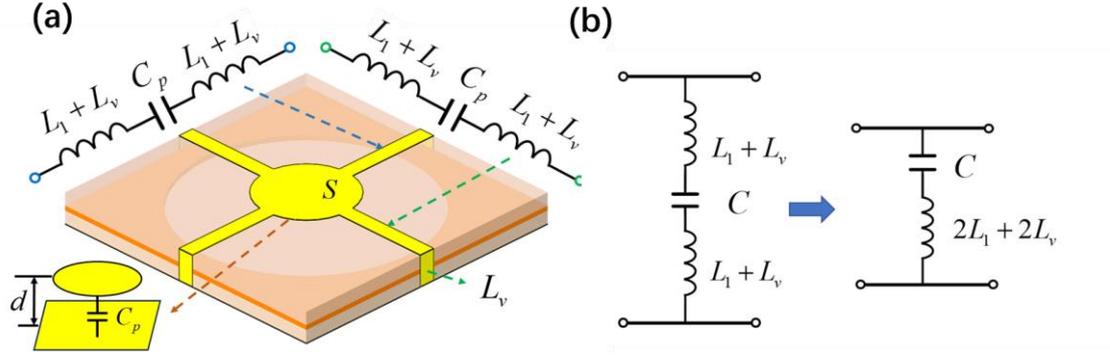

Figure 3 the equivalent circuit of the single patch LC model. (a) The equivalent circuit in TE (blue) and TM (green) mode. (b) The simplified equivalent circuit.

Based on the equivalent circuit model and one-port network theory, the equivalent serial impedances $Z$ of the circuit is expressed as

$$Z = R + jwL_t + \frac{1}{jwc_p} \tag{1}$$

Where $L_t$ represents the total inductor and $C_p$ represents the capacitance of the plate. The capacitance of a parallel plate capacitor can be expressed as:

$$C_p = \frac{\varepsilon_0 \varepsilon_r S}{d} \tag{2}$$

Where $S$ is the area of one of the plates, $d$ is the separation distance between two plates, $\varepsilon_0$ is dielectric constant of the vacuum and $\varepsilon_r$ is the relative dielectric permittivity of the dielectric layer. The change in capacitance can be expressed using calculus and the derivative of the capacitance formula with respect to $d$, which can be expressed by:

$$\frac{dC_p}{dd} = -\frac{\varepsilon_0 \varepsilon_r S}{d^2} \tag{3}$$

This derivative represents the rate of change of capacitance with respect to the separation distance $d$. Thus, supposing $S$ is a constant term and the initial value of the capacitor is $C_p(0)$, as the distance d is changed by a small amount $\Delta d$, we have the following equation:

$$\Delta C = -\frac{\varepsilon_0 \varepsilon_r S}{d^2} \times \Delta d \tag{4}$$

The total capacitance value is:

$$C = C_0 + \Delta C = \frac{\varepsilon_0 \varepsilon_r S}{d} + \int_{d_0}^{d_0 - \Delta d} \left(-\frac{\varepsilon_0 \varepsilon_r S}{d^2}\right) dd = \frac{\varepsilon_0 \varepsilon_r S}{d - \Delta d} \tag{5}$$

The capacitance at zero pressure is given by:

$$C_p(0) = \frac{\pi \varepsilon_0 r^2}{t_a + \left(\frac{2t_m}{\varepsilon_r}\right)} \tag{6}$$

Where $r$ is the area of metal disc on the surface of the upper dielectric layer, $t_a$ is the depth of the cavity and $t_m$ is the thickness of membrane. With pressure loaded, the expression for variable capacitance is obtained as[24]:

$$C(P) = \frac{C_0}{\sqrt{\gamma}} \tanh^{-1}(\sqrt{\gamma}) \tag{7}$$

Where $\gamma$ can be expressed as:

$$\gamma = \frac{2d_0 \varepsilon_r}{\varepsilon_r h + 2t_m} \tag{8}$$

According to Taylor series expansion with omission of higher terms, Eq. 7 can be approximately equal to

$$C(P) \approx C_0 \left(1 + \frac{\gamma}{3}\right) \qquad (9)$$

The resonant frequency is defined as the frequency of a circuit when the values of capacitive impedance and the inductive impedance become equal. To satisfy the condition of resonance, the circuit must be purely resistive. Hence, for a circuit containing resistor, inductor and capacitor, the imaginary part of impedance is zero, which can be expressed as

$$f = \frac{1}{2\pi}\sqrt{\frac{1}{L_t C_p}} = \frac{1}{2\pi r}\sqrt{\frac{t_a + \left(\frac{2t_m}{\varepsilon_r}\right)}{L_t \pi \varepsilon_0}} \qquad (10)$$

According to the reflection coefficient (S11) parameter analysis shown in Fig.2(d), the component exhibits a sole resonant frequency within the range of 8-12GHz, precisely at 11.9GHz, with a relatively modest Q value. Consequently, there is a necessity to enhance the design for improved performance.

➢ **The Mechanical Analysis**

The sealed cavity enclosed by two deflectable diaphragms of our device is a key component to achieve pressure sensitivity. To gain further understanding of the deflection of the mechanical structures for circular shapes in our sensor design, a simplified model is investigated by thin plate theory. As illustrated in Fig. 4, the thickness of the circular film is $t_{film}$, the maximum deformation amount of the film (i.e., deformation amount at the center of the circle) is $d_0$, the radius of the cylinder air cavity is $R$ and the thickness is $t_{gap}$, $E$ and $v$ represent Young's modulus and Poisson's ratio of the materials, respectively. And the model is suspended and placed in a high-pressure environment (pressure $P$) so as to achieve a double layer deflection.

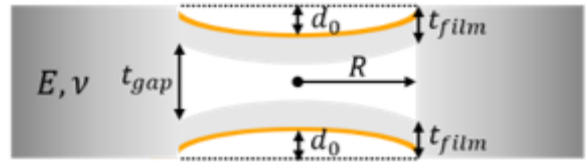

Figure 4 Sensor's electromechanical model under the loaded pressure.

The relationship between the predicted pressure (P) in the environment and the force (F) on the film can be expressed as:

$$F = P \times S \qquad (11)$$

Based on the elastic mechanism, Yang's modulus E of the elastic material can be expressed as:

$$E = \frac{\sigma}{\varepsilon_d} = \frac{F/S}{\Delta d/d} \qquad (12)$$

Firstly, the flexural rigidity of a thin plate defines the bending stiffness when a mechanical load is applied and can be written as[24]

$$D = \frac{E t_{film}^3}{12(1-v^2)} \qquad (13)$$

According to the flexural rigidity, the deformation amount of the deflected surface for either the top or bottom films as a function of radius $r$ can be represented by[25]

$$d(r) = d_0 \left(1 - \frac{r^2}{a^2}\right)^2 \tag{14}$$

Using the energy method, an approximate expression for the center deflection of a uniformly loaded circular plate is derived as[25]

$$d_0 = \frac{PR^4}{64D} \frac{1}{1 + 0.488\left(\frac{d_0}{t_{film}}\right)^2} \tag{15}$$

By using the analytical equation of $d(r)$ Eq. (14) and $C(P)$ Eq. (9), the approximate device deflection distribution and resonant frequency changes with pressure can be obtained, respectively.

## ➢ Pressure sensor Structure Design & fabrication

As a continuation of the previous work, the challenge of designing a passive wireless pressure sensor with enhanced Q factor and ensuring high sensitivity is explored. To realize highly sensitive pressure sensor with wide measurement range, a new-type pressure sensor based on LC resonance embedded on an air cavity is designed and shown in Fig. 5.

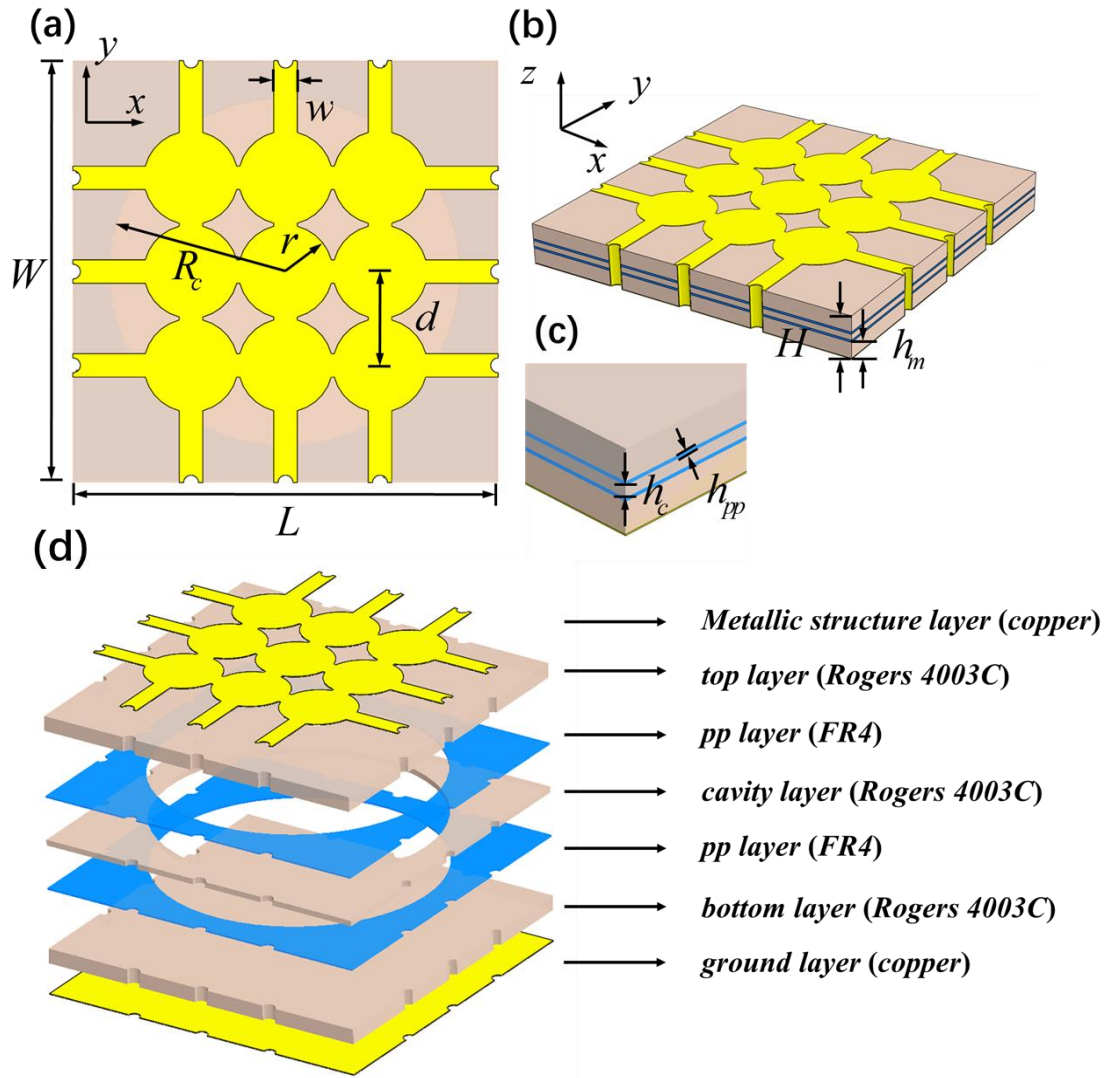

Figure 5 The schematic view of the proposed pressure sensor. (a) Top view. (b) side view. (c) detailed

view. (d) hierarchical structure.

The sensor is mainly composed of 3 layers arranged in a sandwich-like configuration, each of which is the upper plate (0.813mm), the middle air cavity plate (0.305mm) and the lower plate (0.813mm). The middle layer features a void at its center, introducing the cavity essential for sensor functionality. Interbed plates made of FR4 material are placed between each pair of plates to securely bond them together. All three plates are constructed from Rogers 4003C (ep=3.55) due to its material stability and exceptional performance in compression resistance, including notable flexural strength and a high Young's modulus ($19.65 \times 10^9$). The metallic structure on the surface of the top layer and the bottom layer served as the ground are made of copper with the thickness of 0.018mm. Other vital parameters concerning in the design is presented in Table 2.

Table 2 The vital parameters of the proposed pressure sensor

| Parameters | Value (mm) | Parameters | Value (mm) |
|---|---|---|---|
| $r$ | 2.1 | $w$ | 2.1 |
| $H$ | 2.131 | $h_{pp}$ | 0.1 |
| $R_c$ | 7.5 | $d$ | 4.2 |
| $W$ | 18 | $h_m$ | 0.813 |
| $L$ | 18 | $h_c$ | 0.305 |

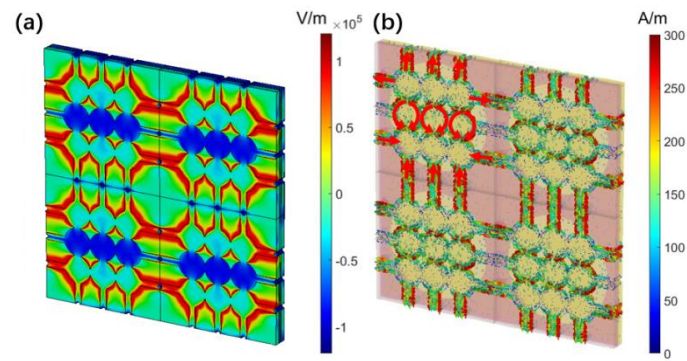

Figure 6 Distribution of the electric field and surface currents at the resonant frequency in TE mode.

The equivalent circuit of the proposed LC pressure sensor, as depicted in Fig.7, can be simplified to three LC series circuits in parallel. It is noteworthy that from Fig. 6(a), the energy is not concentrated around it, and the surface currents flow along its edge. This observation implies that the middle circuit plate functions as an inductor in this design because its extended side length offers a pathway for current flow, rather than acting as a capacitor.

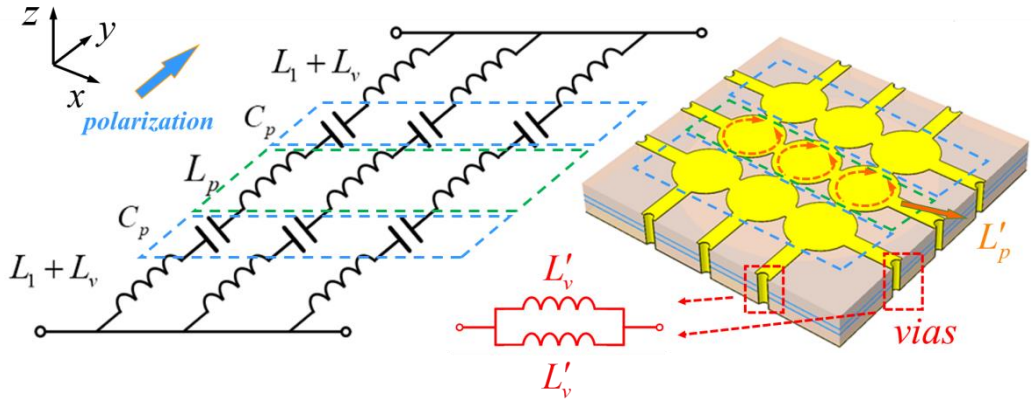

Figure 7 The equivalent circuit of the proposed pressure sensor unit.

## ➢ Mechanical & Electromagnetic Simulation

In the previous section, theoretical calculations for a simplified model that assumes uniform substrate materials are discussed and presented. As for the proposed pressure sensor, which is coated with metallic materials and imbedded with FR-4 interlayers, possess mechanical properties (such as Young's modulus) that differ from those of the substrate. These discrepancies could lead to variations in the deformation levels of the devices. Consequently, to account for these differences and obtain a more accurate representation of the structure's mechanical behavior, we conducted a comprehensive mechanical simulation of the entire structure using COMSOL Multiphysics software.

By setting the corresponding mechanical material parameters of each part of the device in COMSOL, we can obtain the mechanical simulation results of the complete model, as shown in Fig. 8 (b, c, d). For analytical convenience, the model's bottom surface was designated as fixed, and a specific pressure was applied to the top surface. This simulation approach enabled the determination of stress distribution across different sections of the device and the displacement of the materials. Subsequently, we can extract the deformation amounts at the central point of the film under different pressure (Fig. 9(a)). Given the symmetrical thickness of the structure along its vertical axis, it is reasonable to assume that the deformation observed on the bottom membrane mirrors that of the upper surface.

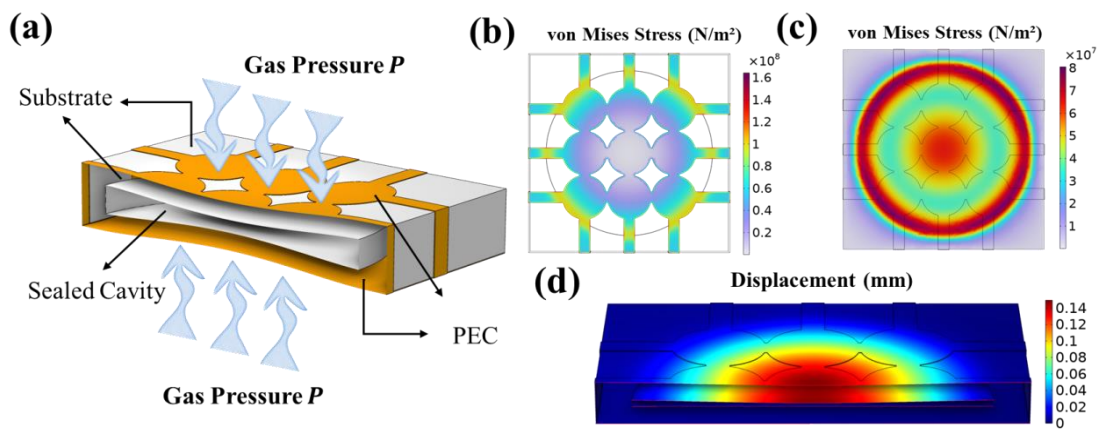

Fig. 8 (a) Sensor model characterization under the loaded pressure and the simplified mechanical model diagram with labeled parameters. (b) von Mises stress distribution in PEC. (c) von Mises stress distribution of the upper surface of the substrate with cavity. (d) Structural displacements in the sensor device under pressure of 1.5MPa.

Furthermore, the resonant frequency decreases as the height of the cavity reduces, as depicted in Fig. 9 with the help of CST Studio. For simplicity, a direct way of changing the height is implemented to see how the frequency changes. Within the limited measurement range, the frequency variations follow a linear trend. This phenomenon primarily arises from alterations in capacitor values. As the distance between the upper and bottom membranes changes, the equivalent capacitance is modified, hence reflecting on the S11 parameter. This result demonstrates the feasibility of the proposed pressure sensor.

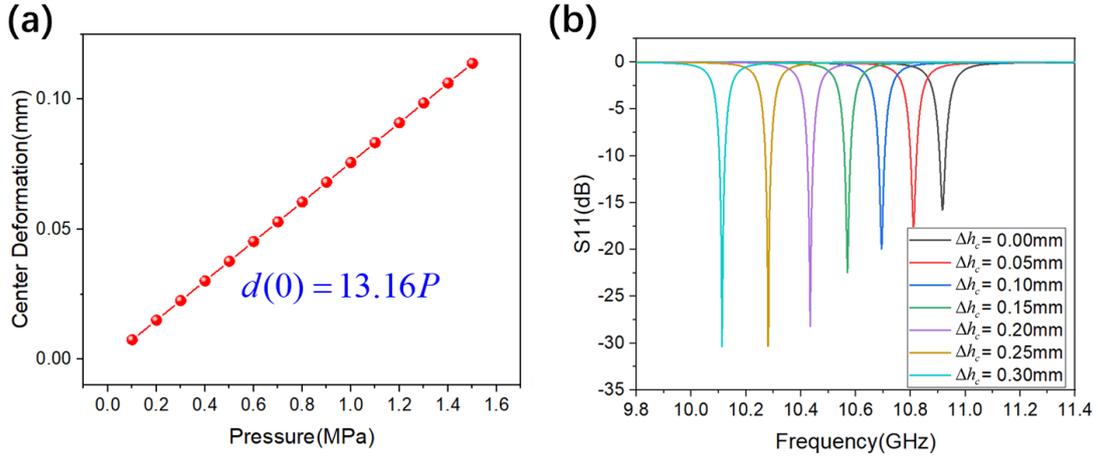

Figure 9 (a) The deformation amounts at the central point of the film under different pressure. (b) The simulated S11 parameter by CST varies as the height of the cavity layer changes from 0.305mm to 0.005mm.

## ➢ Implementation & Measurement

The sensor was positioned on a testing stent within a sealed pressure reactor, while the interrogation antenna was linked to the Agilent E5063A network analyzer using a 50-ohm coaxial cable. The experiment is expanded within the depicted testing setup illustrated in Fig. 10. The Agilent E5063A vector network analyzer (VNA) is connected to the horn antenna which is securely positioned at the top of the pressure reactor via a 50-ohm coaxial cable and adapter. Following calibration, the S11 parameter is measured under standard pressure conditions, as depicted by the pink curve in Fig.11(a), with its resonant frequency observed at 10.932GHz. For the sake of measurement accuracy, the horn antenna aperture is meticulously positioned directly above the sensor, maintaining a distance of 2cm. Additionally, the metal plate with a perforated center beneath the sensor serves multiple functions. Firstly, it acts as a shield against electromagnetic waves by obstructing their passage, thereby minimizing interference. Secondly, it ensures that pressure is evenly applied to both sides of the sensor, promoting consistent and accurate measurements. This strategic placement mitigates potential interference arising from the irregularities of the sealed reactor's metal cavity.

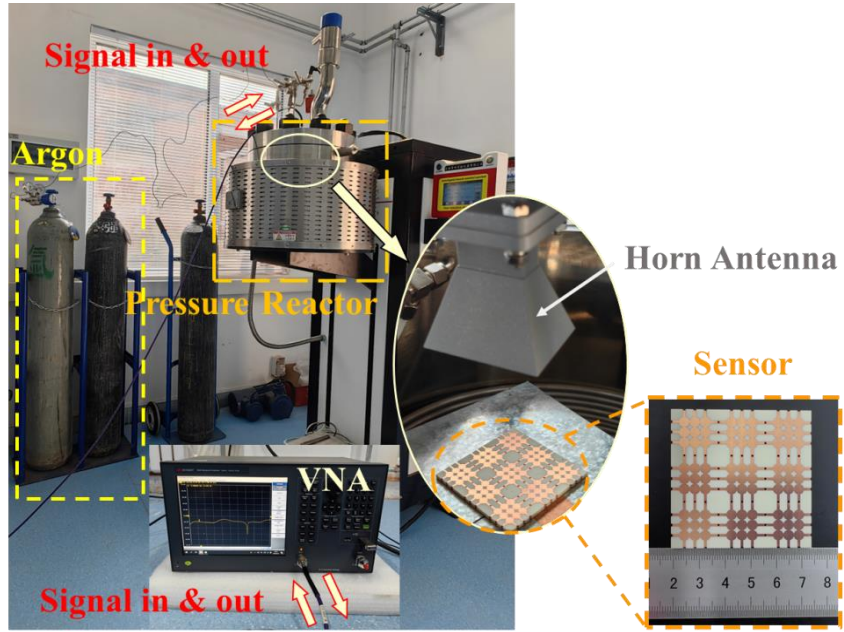

Figure 10 High pressure testing platform.

During the testing process, the argon is loaded into the pressure vessel through a valve and the data curve is recorded at a step of 0.1 MPa. Fig.11(a) shows part of the measured data with the change of pressure from 0 MPa to 1.5 MPa with the step of 0.3 MPa. Due to the decrease of the height of the cavity, when the pressure gets higher, the corresponding frequency gets lower. The experimental frequency of the sensor with no pressure loaded is 10.932 GHz, which slightly deviates from the simulated resonant frequency of 10.917 GHz, mainly because of the processing error during the fabrication of the sensor.

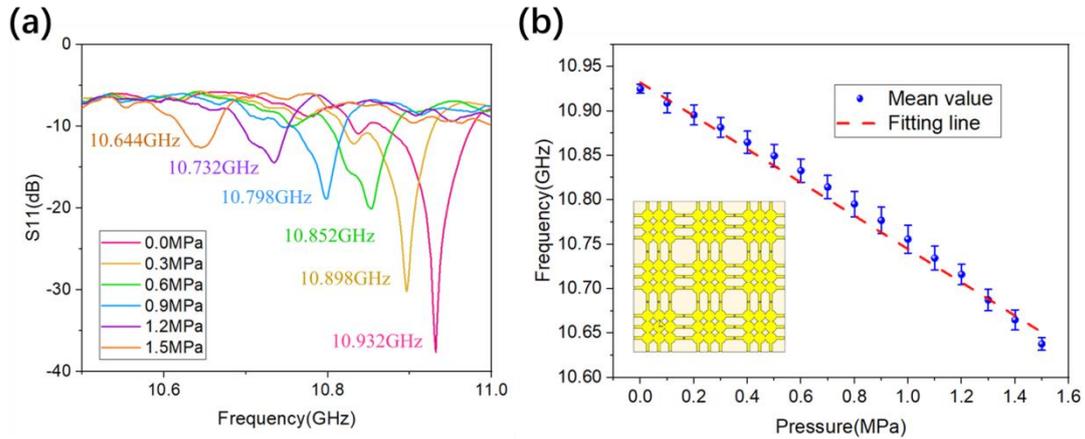

Figure 11 Measurement result. (a) S11 parameter of the sensor under pressure from 0MPa to 1.5MPa. (b) Resonant frequency under the tested pressure from three sets of experiment and the corresponding fitting line.

Fig.11(b) illustrates the relationship between the pressure and the resonant frequency. We conducted three separate sets of experiments at pressure intervals of 0.1MPa. For each pressure point, the average resonant frequency is calculated and plotted against the corresponding pressure value represented by blue dots. The absolute value error for each pressure point, illustrated using error bar colored in bule, is relatively small, with the maximum error reaching 0.01571GHz at 1.0MPa. In Fig.11(b), the fitting curve illustrating the relationship between average resonant frequency ($f$) and pressure ($P$) is depicted by the red dotted line, characterized by the formula

$$f \approx 10.932 - 0.187P \qquad (16)$$

Utilizing the fitting curve, the pressure in the surrounding environment can be determined by measuring the resonance frequency and referencing the corresponding pressure value. This approach enables the proposed pressure sensor to characterize unknown pressures effectively.

## ✧ Discussion

### ➢ Comparison

Table 2 Comparison of the proposed sensor to previous designs

| Reference | Operating Frequency (MHz) | Frequency Shift (MHz) | Pressure range (kPa) | Sensitivity (kHz/kPa) |
|---|---|---|---|---|
| [18] | 70-100 | 4 | 70-120 | 92.98 |
| [26] | 95-103 | 8 | 0-6.7 | 900.36 |
| [27] | 165-185 | 1 | 140-2660(Max) | 3.76 |
| [16] | 30-34 | 2.75 | 0-100 | 26 |
|  | 26-27 | 0.4 | 0-10000 | 0.064 |
| [28] | 14900-15000 | 2.5 | 0-0.9 | 2.5 |
| This work | 8000-12000 | 288 | 0-15000 | 288 |

### ➢ Simulation

To achieve a more precise response curve for enhanced performance understanding and sensor design optimization, a novel approach is proposed for modeling the deflection of the diaphragm in response to air pressure. The conventional method involves simulation by merely adjusting the thickness of the cavity layer to alter the distance between the top and bottom layers. However, this method is overly simplistic as it fails to consider the deformation of the membrane and metallic structure. Consequently, electromagnetic waves are always assumed to be incident vertically, which does not align with practical scenarios. A more concise alternative is proposed herein as a second method. Considering that the deformation of an edge-fixed diaphragm can be equivalently represented as a spherical conformal shape, the modeling involves introducing a sphere to modify the top layer, offering a more streamlined approach, as shown in Figure 12(a).

To quantify the approximation of the deflection of membrane by pressure with spherical conformal model, a common way in mathematic is to calculate the Mean Squared Error (MSE) of the two functions, which can be expressed as

$$MSE = \frac{1}{n}\sum_{i=1}^{n}\bigl(f(x_i) - p(x_i)\bigr)^2 \qquad (17)$$

Where $f(x)$ represents the defined approximation function, and $p(x)$ represents the deflection function of membrane. The MSE value serves as a quantitative measure of the approximation quality between the two functions, with lower MSE values indicating a closer resemblance between them.

By selecting 3001 samples within the predefined domain, the computation of MSE becomes straightforward. For simplification, we exclusively consider the segment associated with the variable and standardize the maximum deformation height. Subsequently, we compute the MSE separately for both the spherical and conventional models, resulting in values after normalization of 0.159 and 0.583 respectively, which indicates that this new one exhibits superior performance compared to the conventional method. Furthermore, we observe that reducing the radius of the sphere leads to a curve that closely aligns with the theoretical value. Fig. 12(b) illustrates the normalized relationship between membrane deflection and cavity radius. In the conventional method, the membrane deflection remains constant as the radius changes, while the proposed method exhibits a trend that closely aligns with theory.

To further enhance the accuracy of the simulation results, a conformal sphere with a radius of 6.5mm is employed in the actual simulation, which is 1mm smaller than the cavity radius. Additionally, to better approximate the real pressure deformation, a bend is incorporated at the joint of the deformation edge. Additionally, we calculate the MSE for the optimized spherical conformal model, depicted as the blue dotted line in the figure. This approach enhances the performance by a factor of 2.26 compared to the unoptimized method.

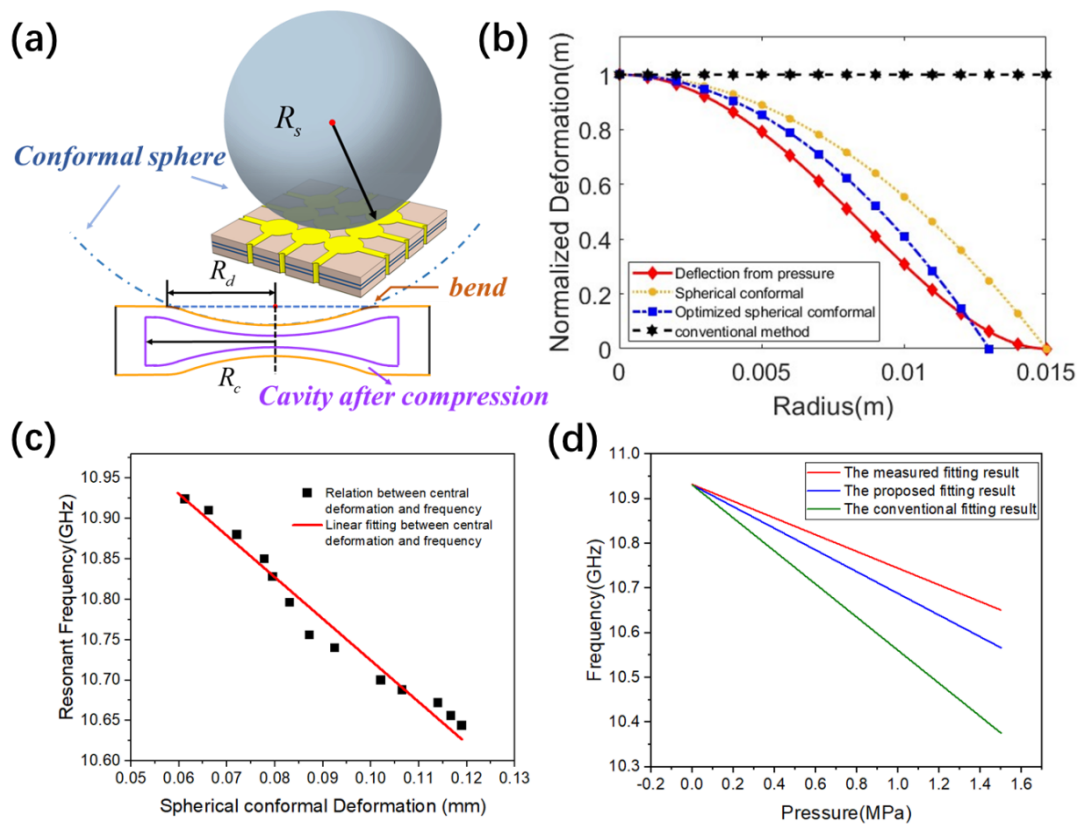

Figure 12 (a) Schematic view of spherical conformal simulation model. (b) The relation between the deformation of the memberane and the radius under pressure. (c) The simulated resonant frequency as the center of membrane deforms at pressure varying from 0MPa to 1.5MPa and the corresponding fitting line. (d) The proposed method and the conventional method compared with the measured fitting result.

Due to the subtle deformation of the membrane, the Time Domain Solver is chosen in CST for a more precise mesh partitioning. The simulated resonant frequency derived from S11 parameter with different deformation is shown in Fig.12(c). Furthermore, the linear fitting operation is performed in this work and the equation is given as

$$f_{sim} = 11.056 - 3.207d \tag{18}$$

Where $d$ represents the deformation of a single membrane. Combined with the relation of deformation at the center of membrane and pressure as depicted in Fig.9(a), the correlation between simulated resonant frequency and pressure can be obtained as

$$f_{sim} = 10.935 - 0.243P \tag{19}$$

In order to accurately compare the performance of the two aforementioned methods, the starting point of the fitting line for both the simulated and measured results is calibrated to the same frequency, as depicted in Fig.12(d). This result highlights their remarkable similarity and suggests that the proposed method can more accurately reflect real measurements to a certain extent.

## ➢ Conclusion

The proposed wireless pressure sensor shows great potential for intrinsically packaged pressure sensors if implemented with higher-temperature materials and embedded circuitry. The pressure test result shows that the resonant frequency of the sensor changed from 10.932GHz to 10.64GHz in the range of 0 to 1.5MPa, and the absolute sensitivity is 192 kHz/kPa. The proposed method of modeling in simulation shows its great accuracy compared to conventional method.